\documentclass[twocolumn,letter]{jpsj3}

\usepackage{color}
\usepackage{bm}

\title{%
Regularization of a Massless Dirac Model to Describe
Anomalous Electromagnetic Response of Weyl Semimetals
}

\author{%
Yositake Takane
}

\inst{%
Department of Quantum Matter, Graduate School of Advanced Sciences of Matter,\\
Hiroshima University, Higashihiroshima, Hiroshima 739-8530, Japan
}

\recdate{ \hspace{50mm} }

\abst{%
An unbounded massless Dirac model with two nondegenerate Dirac cones is
the simplest model for Weyl semimetals,
which show the anomalous electromagnetic response
of chiral magnetic effect (CME) and anomalous Hall effect (AHE).
However, if this model is naively used to analyze the electromagnetic response
within a linear response theory, it gives the result apparently
inconsistent with the persuasive prediction based on a lattice model.
We show that this serious difficulty is related to the breaking of current
conservation in the Dirac model due to quantum anomaly
and can be removed if current and charge operators are redefined
to include the contribution from the anomaly.
We demonstrate that the CME as well as the AHE can be properly described
using newly defined operators,
and clarify that the CME is determined by the competition
between the contribution from the anomaly and that from low-energy electrons.
}

\begin{document}
\sloppy
\maketitle

Weyl semimetals are three-dimensional gapless systems possessing
pairs of nondegenerate Dirac cones with opposite chirality;~\cite{murakami,
wan,yang,burkov1,burkov2,WK,delplace,halasz,sekine}
one Dirac cone with $+$ chirality is separated from the other
with $-$ chirality in energy and/or momentum.
The band touching point of each Dirac cone is called the Weyl node.
A pair of Weyl nodes is robust against weak perturbations
as long as a pair annihilation of Dirac cones does not occur.~\cite{murakami}
The $+$ ($-$) chirality of Weyl nodes corresponds to
the monopole (antimonopole) of Berry curvature with unit strength.
This topological character gives rise to an unusual electromagnetic response,
anomalous Hall effect (AHE),~\cite{burkov1}
and chiral magnetic effect (CME),~\cite{nielsen,zyuzin}
as well as the appearance of protected surface states~\cite{wan}
subjected to spin connection.~\cite{imura}
Thus far, TaAs and NbAs are experimentally identified
as Weyl semimetals.~\cite{xu1,lv1,lv2,xu2}

The CME, i.e., anomalous induction of an electron current in response to
an external magnetic field, has recently attracted significant attention.
Although this arises even in systems with degenerate Dirac cones
if an additional electric field or chiral chemical potential is
applied,~\cite{nielsen,fukushima,son1,son2,basar} we focus on the CME
in Weyl semimetals with nondegenerate Dirac cones
in the presence of only an external magnetic field.~\cite{zyuzin,
liu,vazifeh,chen,goswami,landsteiner,chang,fujita}
Let us assume that a pair of Weyl nodes with $\pm$ chirality is
located at $(\mib{k},E) = (\pm\mib{b},\pm b_{0})$
in momentum and energy space, where $\mib{b}=(b_{1},b_{2},b_{3})$.
For this system, we introduce the following massless Dirac model
with the vector potential $\mib{A}=(A_{1},A_{2},A_{3})$
and the scalar potential $A_{0}$:
\begin{align}
        \label{eq:Hamiltonian}
   H_{\pm} 
 & = \int d \mib{x}
     \psi_{\pm}^{\dagger}(\mib{x})
     \Bigl[ \pm v\mib{\sigma}\cdot(-i\nabla + e\mib{A}(\mib{x}))
            + eA_{0}(\mib{x}) - \mu_{\rm c}
               \nonumber \\
 & \hspace{30mm}
            + v\mib{\sigma}\cdot\mib{b} \pm b_{0}
     \Bigr]
     \psi_{\pm}(\mib{x}) ,
\end{align}
where $\psi_{\pm}(\mib{x})$ represents the spinor field describing
Weyl electrons with $\pm$ chirality, and $v$ and $\mu_{\rm c}$ respectively
denote the velocity and chemical potential.
Here, $\mib{\sigma} = (\sigma_{1},\sigma_{2},\sigma_{3})$ is the Pauli matrix.

Applying a field theoretical technique to this model,
Zyuzin and Burkov~\cite{zyuzin} derived an effective action describing
the electromagnetic response of Weyl semimetals.
This action predicts the AHE in the case of $\mib{b} \neq \mib{0}$,
in accordance with the argument discussed in Ref.~\citen{burkov1}.
It also predicts that, in the case of $b_{0}\neq 0$, an electron current
due to the CME is induced by a magnetic field $\mib{B}$ as
\begin{align}
        \label{eq:ZB-CME}
  \mib{j} = -\frac{e^{2}b_{0}}{2\pi^{2}}\mib{B} .
\end{align}
This indicates that a finite current appears even at equilibrium
in response to a static magnetic field.
Several authors have examined this intriguing prediction by adopting
a lattice model,~\cite{vazifeh,chang} which is free from
the pathological features of the massless Dirac model
with an unbounded energy spectrum.
The following conclusion is established; the CME current vanishes
under a static magnetic field at equilibrium but can appear
in nonequilibrium situations with a time-dependent magnetic field.
However, it is still unclear why the effective action of Ref.~\citen{zyuzin}
partially fails to describe the electromagnetic response.
A reexamination based on the massless Dirac model is desirable
to resolve this problem.

In this Letter, we reconsider the CME as well as the AHE in Weyl semimetals
on the basis of the unbounded massless Dirac model.
We calculate corresponding response functions within a linear response theory
and find that the resulting ones are apparently inconsistent
with those obtained from a lattice model.
Then, we show that this serious difficulty is related to
the breaking of current conservation due to a quantum anomaly,
and that it can be removed if current and charge operators are redefined
to include the contribution from the anomaly.~\cite{liu,landsteiner}
With these operators, the anomalous electromagnetic response can be
properly described using the massless Dirac model.
Once the difficulty is removed, the model has an advantage
of analyzing various properties of Weyl semimetals in a simple manner.
Indeed, we succeeded in obtaining an analytical expression of the coefficient
of the CME, from which we understand that the CME is determined
by the competition between
the contribution from the anomaly and that from low-energy electrons.
We finally present a plausible explanation that partially accounts for
the puzzling nature of the effective action of Ref.~\citen{zyuzin}.
We set $\hbar = k_{\rm B} = 1$ throughout this letter.

We show that, within a linear response theory,
the Dirac model provides an apparently insufficient description
of the anomalous electromagnetic response.
For our argument, it is convenient to decompose the electron current $j^{\mu}$
with $\mu = 1,2,3$ and the charge $j^{0}=\rho$ as
\begin{align}
       \label{eq:def-j-mu}
   j^{\mu} = j_{+}^{\mu} + j_{-}^{\mu} ,
\end{align}
where $j_{\pm}^{\mu}$ represents the contribution from the valley centered at
$\mib{k} = \pm\mib{b}$:
\begin{align}
       \label{eq:def-current}
  j_{\pm}^{\mu}
  = \mp ev \psi_{\pm}^{\dagger}(\mib{x})\sigma_{\mu}\psi_{\pm}(\mib{x}) ,
\end{align}
for $\mu = 1,2,3$ and
\begin{align}
       \label{eq:def-charge}
  j_{\pm}^{0} 
  = \rho_{\pm} = - e \psi_{\pm}^{\dagger}(\mib{x})\psi_{\pm}(\mib{x}) .
\end{align}
Within a linear response theory, the average current and charge induced
by $\mib{A}$ is expressed as
\begin{align}
  \langle j_{\pm}^{\mu} \rangle
 & = -\Pi_{\pm}^{\mu\lambda}(\mib{q},\omega)A_{\lambda}(\mib{q},\omega) ,
      \\
  \langle \rho_{\pm} \rangle
 & = \mp\frac{1}{v}\Pi_{\pm}^{0\lambda}(\mib{q},\omega)
      A_{\lambda}(\mib{q},\omega) .
\end{align}
The response function $\Pi_{\pm}^{\mu\lambda}$ can be obtained by using
the analytic continuation of $i\nu \to \omega+i\delta$
from its Matsubara representation,
\begin{align}
           \label{eq:def-response}
    \Pi_{\pm}^{\mu\lambda}(\mib{q},i\nu)
  &  = \frac{e^{2}v^{2}}{\beta V}\sum_{\mib{k}}\sum_{\epsilon}
          \nonumber \\
  &    \hspace{-8mm} \times
       {\rm Tr} \left\{\sigma_{\mu}G_{\pm}(\mib{k}+\mib{q},i\epsilon+i\nu)
                       \sigma_{\lambda}G_{\pm}(\mib{k},i\epsilon)\right\} ,
\end{align}
where $\beta$ and $V$ are respectively the inverse of the temperature $T$
and the volume of the system.
Here, the thermal Green's function is given by
\begin{align}
  G_{\pm}(\mib{k},i\epsilon)
  = \frac{P_{\pm}(\hat{\mib{k}}_{\pm})}{i\epsilon+\mu_{\pm}-v|\mib{k}_{\pm}|}
  + \frac{P_{\mp}(\hat{\mib{k}}_{\pm})}{i\epsilon+\mu_{\pm}+v|\mib{k}_{\pm}|} ,
\end{align}
where $\mib{k}_{\pm} = \mib{k}\pm \mib{b}$,
$\mu_{\pm} = \mu_{\rm c} \mp b_{0}$, and
\begin{align}
  P_{\pm} = \frac{1}{2}\left(1\pm\mib{\sigma}\cdot\hat{\mib{k}}_{\pm} \right)
\end{align}
with $\hat{\mib{k}}_{\pm} = \mib{k}_{\pm}/|\mib{k}_{\pm}|$.

Let us consider the AHE within the framework presented above.
We calculate $\Pi_{\pm}^{21}$
to obtain the Hall current $\langle j^{2}\rangle$ in the $y$-direction
induced by an electric field $\mib{E}=(E_{1},0,0)$, where we assume
that $\mib{E}(\mib{q},\omega) = i\omega \mib{A}(\mib{q},\omega)$
with $\mib{A}=(A_{1},0,0)$.
The Hall current is expressed as
$\langle j^{2} \rangle = \sigma^{\rm AHE} E_{1}$
with $\sigma^{\rm AHE}$ being the Hall conductivity.
In the uniform limit of $\mib{q} = \mib{0}$,
the Hall conductivity is given by
\begin{align}
    \sigma^{\rm AHE}(\omega)
    = - \frac{1}{i\omega}\Pi^{21}(\mib{0},i\nu)\big|_{i\nu \to \omega+i\delta}
\end{align}
with $\Pi^{21}\equiv\Pi_{+}^{21}+\Pi_{-}^{21}$.
From Eq.~(\ref{eq:def-response}), performing the trace and 
the Matsubara summation, we find
\begin{align}
  \Pi_{\pm}^{21}(\mib{0},i\nu)
  & = \mp i \frac{e^{2}v^{2}}{V}\sum_{\mib{k}}
      \sum_{\eta=\pm}\eta(\hat{\mib{k}}_{\pm})_{z}
           \nonumber \\
  & \hspace{-7mm} \times
      \frac{f_{\rm FD}(\eta v|\mib{k}_{\pm}|-\mu_{\pm})
            - f_{\rm FD}(-\eta v|\mib{k}_{\pm}|-\mu_{\pm})}
           {i\nu + 2 \eta v|\mib{k}_{\pm}|} ,
\end{align}
where $f_{\rm FD}$ denotes the Fermi-Dirac function.
Under the ultraviolet energy cutoff of
$-v|\mib{k}_{\pm}|\pm b_{0} > -E_{\rm c} $,~\cite{basar,landsteiner}
the integral over $\mib{k}$ vanishes
owing to the presence of the factor $(\hat{\mib{k}}_{\pm})_{z}$.
Hence, the Hall conductivity vanishes as
\begin{align}
      \label{eq:AHE1-f}
  \sigma^{\rm AHE} \overset{?}{=} 0 .
\end{align}
This indicates the absence of the AHE in Weyl semimetals, apparently
contradicting the prediction based on a lattice model.~\cite{vazifeh,chang}
As a phenomenon related to the AHE,
let us also consider charge induction in response to a magnetic field.
We calculate $\Pi_{\pm}^{02}$ to obtain
the charge $\langle \rho\rangle$ induced by a magnetic field
$\mib{B}=(0,0,B_{3})$, where we assume that
$\mib{B}(\mib{q},\omega) = i\mib{q}\times\mib{A}(\mib{q},\omega)$
with $\mib{A}=(0,A_{2},0)$ and $\mib{q} = (q,0,0)$.
The charge is expressed as
$\langle \rho \rangle = \alpha^{\rm AHE} B_{3}$, where
\begin{align}
   \alpha^{\rm AHE}(\mib{q},\omega)
   = - \frac{1}{v(iq)}
       \Pi^{02}(\mib{q},i\nu)\big|_{i\nu \to \omega+i\delta}
\end{align}
with $\Pi^{02}\equiv\Pi_{+}^{02}-\Pi_{-}^{02}$.
The response function is given by
\begin{align}
  \Pi_{\pm}^{02}(\mib{q},i\nu)
  & = \pm \frac{e^{2}v^{2}}{2V}\sum_{\mib{k}}\sum_{\eta,\eta'=\pm}
      \left(\eta(\hat{\mib{k}}_{\pm})_{y}+\eta'(\hat{\mib{k}'}_{\pm})_{y}
      \right)
          \nonumber \\
  & \hspace{-10mm} \times
      \frac{f_{\rm FD}(\eta v|\mib{k}_{\pm}|-\mu_{\pm})
            - f_{\rm FD}(-\eta' v|{\mib{k}'}_{\pm}|-\mu_{\pm})}
           {i\nu+\eta v|\mib{k}_{\pm}|-\eta' v|{\mib{k}'}_{\pm}|} ,
\end{align}
where $\mib{k'}_{\pm} = \mib{k}_{\pm}+\mib{q}$.
Under the ultraviolet energy cutoff, the integral again vanishes.
Thus, we find
\begin{align}
      \label{eq:AHE2-f}
  \alpha^{\rm AHE} \overset{?}{=} 0 .
\end{align}
This also contradicts the existing result
based on a lattice model.~\cite{vazifeh}

Let us turn to the CME.
We calculate $\Pi_{\pm}^{32}$ to obtain the CME current $\langle j^{3}\rangle$
in the $z$-direction induced by a magnetic field $\mib{B}=(0,0,B_{3})$,
where $\mib{B}(\mib{q},\omega) = i\mib{q}\times\mib{A}(\mib{q},\omega)$
with $\mib{A}=(0,A_{2},0)$ and $\mib{q} = (q,0,0)$.
The CME current is expressed as
$\langle j^{3} \rangle = \alpha^{\rm CME}B_{3}$, where
\begin{align}
      \label{eq:CME-C}
   \alpha^{\rm CME}(\mib{q},\omega)
   = - \frac{1}{iq}\Pi^{32}(\mib{q},i\nu)\big|_{i\nu \to \omega+i\delta}
\end{align}
with $\Pi^{32}\equiv\Pi_{+}^{32}+\Pi_{-}^{32}$.
The response function is
\begin{align}
  \Pi_{\pm}^{32}(\mib{q},i\nu)
  & = \mp i \frac{e^{2}v^{2}}{2V}\sum_{\mib{k}}\sum_{\eta,\eta'=\pm}
      \left( \eta(\hat{\mib{k}}_{\pm})_{x}
             - \eta'(\hat{\mib{k}'}_{\pm})_{x}\right)
          \nonumber \\
  & \hspace{-7mm} \times
      \frac{f_{\rm FD}(\eta v|\mib{k}_{\pm}-\mu_{\pm}|)
            - f_{\rm FD}(\eta' v|\mib{k'}_{\pm}-\mu_{\pm}|)}
           {i\nu + \eta v|\mib{k}_{\pm}| -\eta'v|\mib{k'}_{\pm}|} .
\end{align}
In calculating $\Pi^{32}$, we separately treat the intraband contribution,
$\Pi_{\rm intra}^{32}$, arising from the terms with $\eta = \eta'$
and the interband contribution, $\Pi_{\rm inter}^{32}$, arising from
the terms with $\eta \neq \eta'$.
In the nearly uniform regime with a small $q$,
their expressions at $T = 0$ are given by
\begin{align}
       \label{eq:Pi-intra}
 & \frac{\Pi_{\rm intra}^{32}}{iq}
   = \frac{e^{2}b_{0}}{3\pi^{2}}
     \left[1-\frac{-i\omega}{\sqrt{(vq)^{2}-(\omega+i\delta)^{2}}}\right] ,
        \\
        \label{eq:Pi-inter}
 & \frac{\Pi_{\rm inter}^{32}}{iq}
   = \frac{e^{2}}{12\pi^{2}}
     \left[ \frac{(b_{0}-\mu_{\rm c})^{2}}
                 {b_{0}-\mu_{\rm c}+ \frac{\omega+i\delta}{2}}
          + \frac{(b_{0}+\mu_{\rm c})^{2}}
                 {b_{0}+\mu_{\rm c}-\frac{\omega+i\delta}{2}}
          - \xi
     \right] ,
\end{align}
where
\begin{align}
   \xi = \left\{ \int^{\epsilon_{1+}}_{\epsilon_{1-}}dE
               - \int^{\epsilon_{2+}}_{\epsilon_{2-}}dE
         \right\}
         \frac{\omega E}{E^{2}-\left(\frac{\omega+i\delta}{2}\right)^{2}}
\end{align}
with $\epsilon_{1\pm} \equiv \max\{\pm b_{0}+\mu_{\rm c},0 \}$
and $\epsilon_{2\pm} \equiv \max\{ \pm b_{0}-\mu_{\rm c},0 \}$.
If $\omega$ is also as small as $2|b_{0}\pm\mu_{\rm c}| \gg \omega$,
the interband contribution is reduced to
\begin{align}
        \label{eq:Pi-inter-lf}
 & \frac{\Pi_{\rm inter}^{32}}{iq}
   = \frac{e^{2}b_{0}}{6\pi^{2}} .
\end{align}
Adding Eqs.~(\ref{eq:Pi-intra}) and (\ref{eq:Pi-inter-lf})
and then substituting it into Eq.~(\ref{eq:CME-C}),
we find that $\alpha^{\rm CME}$ in the nearly static and uniform regime is
\begin{align}
         \label{eq:CME-1}
  \alpha^{\rm CME}
  = -\frac{e^{2}b_{0}}{2\pi^{2}}
     \left[1-\frac{2}{3}
             \frac{-i\omega}{\sqrt{(vq)^{2}-(\omega+i\delta)^{2}}}\right].
\end{align}
In the static limit of $\omega \to 0$ before $q \to 0$,
Eq.~(\ref{eq:CME-1}) gives
\begin{align}
  \alpha^{\rm CME} \overset{?}{=} -\frac{e^{2}b_{0}}{2\pi^{2}} ,
\end{align}
which indicates that the persistent current flows in the direction parallel
to an applied magnetic field.
Again, this conclusion apparently contradicts the prediction based on
a lattice model.~\cite{vazifeh,chang}
Note here that the contribution to $\alpha^{\rm CME}$,
i.e., Eqs.~(\ref{eq:Pi-intra}) and (\ref{eq:Pi-inter}), arises from
only low-energy electrons near the Fermi level.
This is also inconsistent with the argument in Ref.~\citen{chang}.

We resolve the discrepancies mentioned above by relying on the concept of
quantum anomaly in the relativistic field theory.
We show that the current and charge defined
in Eqs.~(\ref{eq:def-j-mu})--(\ref{eq:def-charge}) are inappropriate
in considering the electromagnetic response of Weyl semimetals.
For simplicity, we set $\mu_{\rm c} = 0$ in the argument given below.
Let us start with the partition function of the massless Dirac model:
\begin{align}
  Z = \int D\psi^{\dagger}D\psi \: e^{S} ,
\end{align}
where $S = S_{+} + S_{-}$ with
\begin{align}
        \label{eq:Action}
   S_{\pm} 
 & = \int_{0}^{\beta} d\tau \int d\mib{x} \,
     \psi_{\pm}^{\dagger}(\mib{x},\tau)
     \Bigl[-\partial_{\tau} - eA_{0}
              \nonumber \\
 & \hspace{5mm}
            \mp v\mib{\sigma}\cdot(-i\nabla + e\mib{A})
            - v\mib{\sigma}\cdot\mib{b} \mp b_{0}
     \Bigr]
     \psi_{\pm}(\mib{x},\tau) .
\end{align}
To rewrite $S$ in a relativistic form,
we define the $\gamma$ matrices as~\cite{comment1}
$\gamma^{0} = \tau_{1}\otimes\sigma_{0}$,
$\gamma^{\mu} =-i\tau_{2}\otimes\sigma_{\mu}$ for $\mu = 1,2,3$,
$\gamma^{4} = i\gamma^{0}$, and
$\gamma_{5} = i\gamma^{0}\gamma^{1}\gamma^{2}\gamma^{3}$,
where $\sigma_{0}={\rm diag}(1,1)$,
and $\tau_{1}$ and $\tau_{2}$ are the first and second components
of the Pauli matrix
representing the valley degrees of freedom, respectively.
We also define the projection operators as
\begin{align}
   P_{L} = \frac{1-\gamma_{5}}{2} , \hspace{5mm}
   P_{R} = \frac{1+\gamma_{5}}{2} .
\end{align}
Let us introduce four component electron fields $\psi$ and $\bar{\psi}$,
\begin{align}
  \psi = {}^t\! (\psi_{+},\psi_{-}) ,
  \hspace{6mm}
  \bar{\psi} = \psi^{\dagger}\gamma^{0}
             = (\psi_{-}^{\dagger}, \psi_{+}^{\dagger}) .
\end{align}
With these notations and $x_{4}=\tau$, $A_{4}=-iA_{0}$, and $b_{4}=-ib_{0}$,
the action is compactly written as~\cite{fujikawa}
\begin{align}
        \label{eq:Action_rel}
   S
 & = \int d^{4}x \,
     \bar{\psi} i\gamma^{\mu}
     \Bigl(\partial_{\mu} + ieA_{\mu}^{L}P_{L} + ieA_{\mu}^{R}P_{R}
     \Bigr) \psi ,
\end{align}
where $v$ is eliminated by a rescaling,~\cite{comment2} and
\begin{align}
  A_{\mu}^{L} = A_{\mu} - \frac{b_{\mu}}{e} ,\hspace{5mm}
  A_{\mu}^{R} = A_{\mu} + \frac{b_{\mu}}{e} .
\end{align}
Hereafter, we treat $b_{\mu}$ as a fluctuating gauge field.
The relevant result for Weyl semimetals is obtained by taking
the static and uniform limit with respect to $b_{\mu}$.

As is well known in the relativistic field theory,
$\langle\bar{\psi}\gamma^{\mu}\psi\rangle$ obeys
the following relation:~\cite{fujikawa}
\begin{align}
        \label{eq:covariant-ano1}
  \partial_{\mu}\langle\bar{\psi}\gamma^{\mu}\psi\rangle
   = - \frac{ie^{2}}{32\pi^{2}}\varepsilon^{\mu\lambda\rho\sigma}
       \left(F_{\mu\lambda}^{L}F_{\rho\sigma}^{L}
             - F_{\mu\lambda}^{R}F_{\rho\sigma}^{R}\right) ,
\end{align}
where $\varepsilon^{1234} = 1$ and
\begin{align}
  F_{\mu\lambda}^{L,R}
  = \partial_{\mu}A_{\lambda}^{L,R} - \partial_{\lambda}A_{\mu}^{L,R} .
\end{align}
The right-hand side of Eq.~(\ref{eq:covariant-ano1}) is called
a covariant anomaly and is regarded as the contribution from
the ultraviolet regime of $E \to -\infty$.
It is convenient to modify Eq.~(\ref{eq:covariant-ano1}) to
\begin{align}
          \label{eq:covariant-ano2}
    \partial_{\mu}\langle\bar{\psi}\gamma^{\mu}\psi\rangle
   = \frac{ie}{8\pi^{2}}\varepsilon^{\mu\lambda\rho\sigma}
     f_{\mu\lambda}F_{\rho\sigma}
\end{align}
with $f_{\mu\lambda}=\partial_{\mu}b_{\lambda}-\partial_{\lambda}b_{\mu}$ and
$F_{\mu\lambda}=\partial_{\mu}A_{\lambda}-\partial_{\lambda}A_{\mu}$.
Let us move to the real-time representation by performing the Wick rotation of
$\tau \to i t$ with $A_{4} \to -iA_{0}$ and $b_{4} \to -ib_{0}$
and restore $v$ by returning to the original scale.~\cite{comment2}
Noting Eqs.~(\ref{eq:def-j-mu})--(\ref{eq:def-charge}), we find that
Eq.~(\ref{eq:covariant-ano2}) is modified to
\begin{align}
        \label{eq:CCL-breaking}
  \partial_{\mu}\langle j^{\mu}\rangle
  = - \frac{e^{2}}{8\pi^{2}}\varepsilon^{\mu\lambda\rho\sigma}
      f_{\mu\lambda}F_{\rho\sigma} 
  = - \partial_{\mu} j_{\rm A}^{\mu} ,
\end{align}
where $\varepsilon^{1230} = 1$ and
\begin{align}
          \label{eq:A-current}
   j_{\rm A}^{\mu}
   = \frac{e^{2}}{2\pi^{2}}\varepsilon^{\mu\lambda\rho\sigma}
     b_{\lambda}\partial_{\rho}A_{\sigma} .
\end{align}
This clearly indicates the breaking of current conservation,
so the naive definition of $j^{\mu}$ given
in Eqs.~(\ref{eq:def-j-mu})--(\ref{eq:def-charge})
cannot be justified in the context of condensed matter physics.
A reasonable way to avoid this difficulty is
to redefine $j^{\mu}$ as~\cite{liu,landsteiner}
\begin{align}
       \label{eq:new-j}
   j^{\mu} = -ev\left(\psi_{+}^{\dagger}\sigma_{\mu}\psi_{+}
                 - \psi_{-}^{\dagger}\sigma_{\mu}\psi_{-}
             \right) + j_{\rm A}^{\mu}
\end{align}
for $\mu = 1,2,3$ and 
\begin{align}
       \label{eq:new-rho}
   j^{0} = \rho = -e\left(\psi_{+}^{\dagger}\psi_{+}
                        + \psi_{-}^{\dagger}\psi_{-}
                    \right) + j_{\rm A}^{0} .
\end{align}
Clearly, the newly defined $j^{\mu}$ obeys the ordinary conservation law,
i.e., $\partial_{\mu}\langle j^{\mu}\rangle = 0$.
A brief comment on the above argument is in order here.
If the static and uniform limit is taken for $b_{\mu}$, the central part of
Eq.~(\ref{eq:CCL-breaking}) apparently vanishes as $f_{\mu\lambda}=0$
and hence the current conservation is not broken.
However, even in this limit, $j_{\rm A}^{\mu}$ does not vanish and
the definition of $j^{\mu}$ in Eqs.~(\ref{eq:new-j}) and (\ref{eq:new-rho})
should be justified to guarantee the continuity of the field theory
with respect to $b_{\mu}$.

Let us briefly consider the meaning of $j_{\rm A}^{\mu}$.
In lattice models with a bounded spectrum, the contribution to
electron current generally consists of two components:
one from low-energy electron states closely near the Fermi level
and the other from filled Fermi sea
including electron states far away from the Fermi level.
Apparently, $j_{\rm A}^{\mu}$ corresponds to the latter,
representing a thermodynamic contribution.
Our argument suggests that, in the Dirac model with an unbounded spectrum,
such a bulk contribution is accumulated into the ultraviolet regime
and manifests itself in the form of anomalies.
Although our derivation of $j_{\rm A}^{\mu}$ assumes $\mu_{\rm c} = 0$
and $T = 0$,~\cite{comment3}
$j_{\rm A}^{\mu}$ should not depend on $\mu_{\rm c}$ and $T$
as it arises from the ultraviolet regime of $E \to -\infty$.

Now, we show that, with the newly defined expression of $j^{\mu}$,
we can properly describe the anomalous electromagnetic response
in Weyl semimetals by using the unbounded massless Dirac model.
As $j_{\rm A}^{\mu}$ corresponds to the bulk contribution
from the filled Fermi sea,
the correct response is obtained by adding $j_{\rm A}^{\mu}$ to
the linear-response contribution arising from low-energy electrons.
Let us consider the AHE in the case of $\mib{b} = (0,0,b_{3})$
focusing on the Hall current in the $y$-direction
under an electric field $E_{1}$ in the $x$-direction.
In this case, Eq.~(\ref{eq:A-current}) results in
\begin{align}
  j_{\rm A}^{2} = \frac{e^{2}b_{3}}{2\pi^{2}}E_{1} .
\end{align}
Since the linear-response contribution vanishes,
the Hall current is solely determined by the anomaly.
Hence, the Hall conductivity is given by
\begin{align}
  \sigma^{\rm AHE} = \frac{e^{2}b_{3}}{2\pi^{2}} .
\end{align}
Also, the charge response under a magnetic field $B_{3}$ in the $z$-direction
is solely determined by the anomaly as
\begin{align}
  j_{\rm A}^{0} = -\frac{e^{2}b_{3}}{2\pi^{2}}B_{3} .
\end{align}
Hence, the corresponding coefficient is
\begin{align}
  \alpha^{\rm AHE} = -\frac{e^{2}b_{3}}{2\pi^{2}} .
\end{align}

Let us turn to the CME in the case of $b_{0} \neq 0$
under a magnetic field $B_{3}$ in the $z$-direction.
In this case, Eq.~(\ref{eq:A-current}) results in
\begin{align}
          \label{eq:j-3_A}
  j_{\rm A}^{3} = \frac{e^{2}b_{0}}{2\pi^{2}}B_{3} .
\end{align}
In the nearly static and uniform regime, adding the linear-response
contribution, Eq.~(\ref{eq:CME-1}), to the anomaly contribution,
we obtain the coefficient of the CME as
\begin{align}
  \alpha^{\rm CME}
  = \frac{e^{2}b_{0}}{3\pi^{2}}
    \frac{-i\omega}{\sqrt{(vq)^{2}-(\omega+i\delta)^{2}}} .
\end{align}
In the static limit of $\omega \to 0$ before $q \to 0$, we find
\begin{align}
   \alpha^{\rm CME} = 0 ,
\end{align}
in agreement with the prediction based on a lattice model.~\cite{vazifeh,chang}
This indicates that, at equilibrium, the contribution from the anomaly is
exactly screened by low-energy electrons.
Let us next consider the nonequilibrium situation in the presence of
a time-dependent external magnetic field.
In the limit of $q \to 0$ before $\omega \to 0$,
the coefficient of the CME becomes
\begin{align}
         \label{eq:alpha_CME-LF}
  \alpha^{\rm CME}
  = \frac{e^{2}b_{0}}{3\pi^{2}} .
\end{align}
As stressed in Ref.~\citen{chen}, this limit corresponds to the low-frequency
limit of the response to a time-dependent external magnetic field.
Equation~(\ref{eq:alpha_CME-LF}) indicates that, in this case,
low-energy electrons only partially screen the anomaly contribution.
The above expression is justified
as long as $2|b_{0}\pm\mu_{\rm c}|\gg \omega$.
We next consider the high-frequency limit
of $\omega \gg 2|b_{0}\pm\mu_{\rm c}|$.
Note that the anomaly contribution does not depend on $\omega$.
For the linear response contribution, the intraband part vanishes
as can be seen from Eq.~(\ref{eq:Pi-intra}),
while the interband part is obtained from Eq.~(\ref{eq:Pi-inter}) as
\begin{align}
  \frac{\Pi_{\rm inter}^{32}}{iq}
   = -\frac{2e^{2}b_{0}}{3\pi^{2}}
      \frac{(b_{0}^{2}+3\mu_{\rm c}^{2})}{\omega^{2}} .
\end{align}
Consequently, $\alpha^{\rm CME}$ is given by
\begin{align}
  \alpha^{\rm CME}
  = \frac{e^{2}b_{0}}{2\pi^{2}}
    \left[1+\frac{4(b_{0}^{2}+3\mu_{\rm c}^{2})}{3\omega^{2}} \right]
\end{align}
for $\omega \gg 2|b_{0}\pm\mu_{\rm c}|$.
Hence, with increasing $\omega$, $\alpha^{\rm CME}$
approaches the limiting value of $e^{2}b_{0}/(2\pi^{2})$,
which is just the contribution from the anomaly.

Finally, we consider why the effective action derived
by Zyuzin and Burkov~\cite{zyuzin} partially fails to describe
the anomalous electromagnetic response of Weyl semimetals.
Starting from the partition function for the massless Dirac model
represented in a path integral form, they eliminate $b_{\mu}$ in terms of
a gauge transformation and obtain the effective action
by adapting Fujikawa's method.~\cite{fujikawa}
The resulting action consists of only the bulk contribution
from the ultraviolet regime,
implying that low-energy electrons play no role in the anomalous response.
This apparently contradicts our linear-response analysis,
which reveals that low-energy electrons crucially affect the CME.
The partial failure of the action should be attributed to the lack of
a low-energy contribution.~\cite{comment4}
Indeed, with this observation, the puzzling nature of the effective action
is partially resolved;~\cite{comment5}
the action can describe the AHE consisting of
only the bulk contribution, while it fails to fully describe the CME
since the CME consists of not only the bulk contribution
but also the low-energy contribution.

\section*{Acknowledgment}

This work was supported by JSPS KAKENHI Grant Number 15K05130.


\begin{thebibliography}{99}
\bibitem{comment1} Our definition of the $\gamma$ matrices
is different from that of Ref.~\citen{zyuzin}.
\bibitem{comment2} In deriving Eq.~(\ref{eq:Action_rel}),
we eliminate $v$ by the rescaling:
$x_{\mu} \to vx_{\mu}$, $A_{\mu} \to v^{-1}A_{\mu}$,
$b_{\mu} \to v^{-1}b_{\mu}$ for $\mu = 1,2,3$, and $\psi \to v^{-3/2}\psi$.
At the derivation of Eq.~(\ref{eq:CCL-breaking}),
we restore $v$ by returning to the original scale.
\bibitem{comment3} Note that $T=0$ is implicitly assumed since we take
the limit of $\beta \to +\infty$ in rewriting $S$
in the form of Eq.~(\ref{eq:Action_rel}).
\bibitem{comment4} A pair of nondegenerate Weyl nodes are placed at
$E = \pm b_{0}$ across $\mu_{\rm c} = 0$ in the original model while
they merge together at $\mu_{\rm c} = 0$ after a gauge transformation.
Such a marked change in the distribution of low-energy electrons
is not at all reflected in the effective action.
This shortcoming is probably caused by the gauge transformation that breaks
the antiperiodicity of $\psi_{\pm}$,
i.e., $\psi_{\pm}(\mib{x},\tau) = -\psi_{\pm}(\mib{x},\tau+\beta)$.
\end{thebibliography}
\end{document}